# Accurate synthesis of Dysarthric Speech for ASR data augmentation

*Mohammad Soleymanpour*[a], *Michael T. Johnson*[a], *Rahim Soleymanpour*[b], *Jeffrey Berry*[c]

[a] Electrical and Computer Engineering, University of Kentucky, Lexington, KY USA 40506
[b] Department of Biomedical Engineering, University of Connecticut, Storrs, CT USA 06269,
[c] Speech Pathology and Audiology, Marquette University, Milwaukee, WI USA 53201

`{m.soleymanpour, mike.johnson}@uky.edu` , `rahim.soleymanpour@uconn.edu`,
`jeffrey.berry@marquette.edu`

**Abstract**

Dysarthria is a motor speech disorder often characterized by reduced speech intelligibility through slow, uncoordinated control of speech production muscles. Automatic Speech recognition (ASR) systems can help dysarthric talkers communicate more effectively. However, robust dysarthria-specific ASR requires a significant amount of training speech, which is not readily available for dysarthric talkers.

This paper presents a new dysarthric speech synthesis method for the purpose of ASR training data augmentation. Differences in prosodic and acoustic characteristics of dysarthric spontaneous speech at varying severity levels are important components for dysarthric speech modeling, synthesis, and augmentation. For dysarthric speech synthesis, a modified neural multi-talker TTS is implemented by adding a dysarthria severity level coefficient and a pause insertion model to synthesize dysarthric speech for varying severity levels.

To evaluate the effectiveness for synthesis of training data for ASR, dysarthria-specific speech recognition was used. Results show that a DNN-HMM model trained on additional synthetic dysarthric speech achieves WER improvement of 12.2% compared to the baseline, and that the addition of the severity level and pause insertion controls decrease WER by 6.5%, showing the effectiveness of adding these parameters. Overall results on the TORGO database demonstrate that using dysarthric synthetic speech to increase the amount of dysarthric-patterned speech for training has significant impact on the dysarthric ASR systems. In addition, we have conducted a subjective evaluation to evaluate the dysarthric-ness and similarity of synthesized speech. Our subjective evaluation shows that the perceived dysartrhic-ness of synthesized speech is similar to that of true dysarthric speech, especially for higher levels of dysarthria. Audio samples are available at https://mohammadelc.github.io/SpeechGroupUKY/

*Keywords:* Dysarthria, speech recognition, Speech-To-Text, Synthesized speech, Data augmentation

## 1. Introduction

Dysarthria is a motor speech disorder, often caused by traumatic injury or neurological dysfunction, that decreases speech intelligibility through slow or uncoordinated control of speech production muscles(Duffy 2019). People with moderate and severe levels of dysarthria may be less able to communicate with others through speech due to poor intelligibility (Mitchell, Bowen et al. 1996). Although individuals with dysarthria may have the cognitive and language abilities to formulate communication, they may not be able to reliably plan and execute the muscle control needed for sufficiently intelligible speech.

To provide dysarthric talkers with better communication or better tools for diagnosis and treatment, speech technologies can be effective. Technologies such as Automatic Speech Recognition (ASR) have the potential to significantly increase the quality of dysarthric speakers' communication. The use of ASR is now widespread, with systems such as Siri, Alexa, and Google assistant in common use. Although these systems work reasonably well with typical speech, they have difficulty understanding dysarthric speech. Having a dysarthria-specific ASR can potentially help dysarthric talkers to be understood better and ameliorate their communication struggles. Different methods have been used to increase the performance of such systems for dysarthric speech, allowing dysarthric individuals to have a robust and reliable aids for communication and improving quality of life.

However, to have reliable and robust dysarthria-related speech applications, there is an essential need to have access to a substantial amount of recorded dysarthric speech for training. Current datasets containing dysarthric speech are insufficient for automatic speech recognition, severity assessment and dysarthric speech intelligibility enhancement tasks. There are a few publicly available dysarthric speech datasets, including TORGO (Rudzicz, Namasivayam et al. 2012), UASpeech (Kim, Hasegawa-Johnson et al. 2008) and Nemours (Menendez-Pidal, Polikoff et al. 1996), each of these have significant limitations in both size and diversity. TORGO is a popular dysarthric speech dataset of aligned acoustic and articulatory recordings from 15 speakers with eight dysarthric speakers (Rudzicz, Namasivayam et al. 2012). Most of the utterances from these datasets consist of single words which do not capture crossword co-articulation or allow for accurate modeling of prosody and pause characteristics in continuous dysarthric speech. Because there are not an adequate number of conversational sentences, ASR systems trained with these types of datasets are often less robust. Modern ASR methods assume that training data includes a sufficiently large set of speakers, often hundreds to thousands of hours of speech data, to adequately capture enough inter-speaker variability. For example, the LibriSpeech and TED_LIUM datasets used for ASR training contain about 1000 and 450 hours of data, respectively, hundreds of times more data than the dysarthric datasets described above.

To address the data insufficiency issues described above for these and other dysarthric speech technology applications, this paper focuses on the development of data augmentation approaches for dysarthric speech applications, mainly for Speech recognition. The core idea proposed here is a combination of domain-based and deep-learning based speech synthesis models that are able to generate accurate speech with variability across the dimensions most important to dysarthric speech technologies, including speaking styles as well as prosodic characteristics like speaking rate and intonation patterns, and pause models that correlate with dysarthric severity level.

*1.1. Related Works*

Data augmentation is a machine learning technique to generate additional supplemental training data. Augmentation has been widely applied to many different domains, including both image and speech processing. For speech applications, augmentation methods have been used to improve speech recognition (Sharma and Hasegawa-Johnson 2013, Cui, Goel et al. 2015, Rabiee, P., and Hoagg, J. B. 2023,Ko, Peddinti et al. 2015, Ko, Peddinti et al. 2017, Rebai, BenAyed et al. 2017, Park, Chan et al. 2019), clinical speech applications (Geng, Xie et al. , Jiao, Tu et al. 2018, Vachhani, Bhat et al. 2018, Mirheidari, Pan et al. 2020), voice scene classification (Han and Lee 2016, Mun, Park et al. 2017, Zhang, Zhang et al. 2018), children's speech technologies (Fainberg, Bell et al. 2016, Nagano, Fukuda et al. 2019, Sheng, Yang et al. 2019, Kathania, Singh et al. 2020), and speaker identification and verification (Cai, Cai et al. 2017, Qin, Cai et al. 2019, Rituerto-González, Mínguez-Sánchez et al. 2019, Wu, Wang et al. 2019). Techniques such as Vocal Tract Length Perturbation (VTLP) and Statistical Feature Mapping approaches have been implemented for Deep Neural Networks (DNN) and Convolutional Neural Networks (CNN) for acoustic modeling (Cui, Goel et al. 2015). For ASR data augmentation, there have been successful methods for tasks such as simulated Room Impulse Resonances (RIRs) (Ko, Peddinti et al. 2017), adding source-point noises (Ko, Peddinti et al. 2017), voice conversion data augmentation (Wang, Kim et al. 2019, Shahnawazuddin, Adiga et al. 2020), and pitch shifting and speech perturbation (Wang, Kim et al. 2019).

For dysarthric research, temporal and speed modification have been applied on normal speech to simulate artificially dysarthric speech (Vachhani, Bhat et al. 2018) and there has also been augmentation work using transformation methods to convert healthy speech to dysarthric speech(Jiao, Tu et al. 2018).

*1.2. Main contribution of this work*

The approach proposed in this paper focuses on the synthesis of dysarthric speech using neural multi-talker speech synthesis. To synthesize dysarthric speech, there is a need to build a system controlling different characteristics of dysarthric speech for generating variant dysarthric speech. As will be discussed later in this paper, and according to a number of studies (Rudzicz, Namasivayam et al. 2012, Zhang, Dang et al. 2014, Bigi, Klessa et al. 2015, Kuo and Tjaden 2016, Yunusova, Graham et al. 2016), such a system should have the following capabilities in order to support generation of authentic and diverse speech: 1) ability to control the speaking rate (duration), pitch, energy for a variety of dysarthria severity levels, 2) ability to learn and model pause behavior of dysarthric speakers (e.g., duration of pause and pause occurrence) and control pause insertion locations and durations 3) ability to learn and model individual voice characteristics of speakers and use these to generate new speaking styles 4) ability to learn and model these characteristics from a small amount of dysarthric speech data.

**2. Methodology**

Recent progress in end-to-end TTS systems such as Tacotron (Wang, Skerry-Ryan et al. 2017, Shen, Pang et al. 2018), FastSpeech (Ren, Hu et al. 2020, Chien, Lin et al. 2021), Deep-Voice (Ping, Peng et al. 2017) support synthesized speech with high quality and naturalness with varying prosody. These improvements in synthesizing speech inspired us to attempt synthesis of realistic dysarthric speech for ASR training data augmentation. Such neural speech synthesizers have been used to generate new utterances for ASR application for low resource languages (Li, Gadde et al. 2018, Mimura, Ueno et al. 2018, Rosenberg, Zhang et al. 2019, Chen, Rosenberg et al. 2020, Rossenbach, Zeyer et al. 2020). Multi-speaker speech synthesis systems can learn prosody characteristics, speaker and style variation extracted from the training set, and can use speaker embeddings to generate speech in a variety of speaker styles (Li, Gadde et al. 2018, Rosenberg, Zhang et al. 2019, Chen, Rosenberg et al. 2020). This allows for generation of relatively large amounts of the high-quality synthesized speech across a range of speaker characteristics and speaking styles.

In this paper, we propose a method based on multi-talker neural TTS to synthesize dysarthric speech to enhance the results of dysarthric ASR. In addition to traditional prosodic variables such as speech rate, energy, and pitch, we add two new variables to control dysarthric severity and extent of pause insertion. These parameters enable us to generate a broad range of synthesized speech to improve the training of dysarthric ASR systems. To assess the effectiveness of the synthetic speech, we evaluate the Deep Neural Network-Hidden Markov Model (DNN-HMM) models with and without augmented speech. Experiments are carried out using the proposed approach on the TORGO dataset.

*2.1. TORGO Dataset*

There are a few publicly available dysarthric speech datasets, including TORGO (Rudzicz, Namasivayam et al. 2012), UASpeech (Joy and Umesh 2018) and Nemours (Menendez-Pidal, Polikoff et al. 1996). These are mainly used to analyze dysarthric speech and to understand the difference between dysarthric and typical speech. TORGO is a popular dysarthric speech database of aligned acoustic and articulatory recordings from 15 speakers, containing 8 dysarthric speakers and 7 controls. This dataset includes non-word, short words, restricted and non-restricted sentences. Two types of microphones were used to record the data, an 8-element microphone array and a head-mounted microphone. The number of utterances for each dysarthric talker averages 700; whereas for normal speakers the average is 1560 (Joy and Umesh 2018).

In TORGO, Dysarthric speakers are categorized into four dysarthria severity levels, Normal, Very Low, Low, and Medium. Table 2.1 shows the Speakers' level of dysarthria severity and their corresponding intelligibility categories. However, because the TORGO Low category includes only a single speaker, we have reformulated these into three categories: Normal, Low Dysarthria (corresponding to TORGO Very Low and TORGO Low), and Moderate Dysarthria(Soleymanpour, Johnson et al. 2021). The standardized Frenchay Dysarthria Assessment was also used to assess the motor functions of each subject, for additional information about the subjects.

Table 1: Properties of various participants in TORGO dataset

| Severity Level | Speaker ID | Number of Utterances | Intelligibility Category |
|---|---|---|---|
| **Normal** | FC01 | 296 | Intelligible |
|  | FC02 | 2183 |  |
|  | FC03 | 1924 |  |
|  | MC01 | 2141 |  |
|  | MC02 | 1112 |  |
|  | MC03 | 1661 |  |
|  | MC04 | 1614 |  |
| **Very low** | F04 | 675 |  |
|  | M03 | 806 |  |
| **Low** | F03 | 1097 | Unintelligible |
| **Medium** | M05(L/M) | 610 |  |
|  | F01 | 228 |  |
|  | M01 | 739 |  |
|  | M02 | 772 |  |
|  | M04 | 659 |  |

*2.2. Synthesis architecture*

For the baseline synthesis model, we modified FastSpeech2 (Ren, Hu et al. 2020) and a recent variant (Chien, Lin et al. 2021)to synthesize dysarthric speech. Figure 1 shows the main block diagram of the proposed method. In the modified version of the FastSpeech2, the energy, pitch and forced-alignment duration (McAuliffe, Socolof et al. 2017) of each speaker's utterances are incorporated into the phoneme hidden sequence through a "variance adaptor" module, resulting in more controllability of these prosodic parameters. By changing coefficients of the pitch, duration and energy predictors in the Variance Adaptor module across a continuous range between 0 and 2, the model generates synthesized speech corresponded to these coefficients during inference.

The multi-talker variant of the FastSpeech2 decoder works like a voice conversion system, making it a multi-talkers TTS (Chien, Lin et al. 2021) capable of generating speech in a wide range of speaking styles. This is a useful capability for speech synthesis for data augmentation because it allows generation of a robust set of training data.

The prosodic characteristics of dysarthric speech greatly differs from typical speech, specifically at moderate and high severity levels. One significant difference between dysarthric and typical speech is that the speaking rate is often substantially slower for talkers with dysarthria (Kent 1996, Mitchell, Bowen et al. 1996). However, this reduced speaking rate is often not consistent throughout the utterance. In addition, dysarthric vocal excitation may be unstable because many individuals with dysarthria cannot effectively control vocal fold closure and vibration. This may cause inconsistent vocal quality and pitch throughout an utterance.

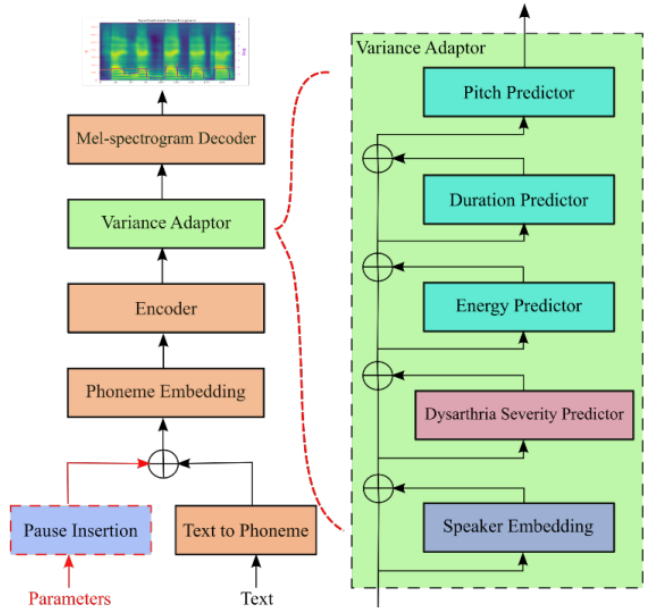

Figure 1. An overview of the proposed architecture

Differences in speech style and speaking rate significantly depend on the dysarthria severity level of the talkers (Soleymanpour, Johnson et al. 2021). To be able to synthesize accurate dysarthric speech, we add a dysarthria severity predictor in the variance adaptor to simulate the characteristics of different severity levels of dysarthric speech. The current version of the model can synthesize dysarthric speech at three different severity levels with discrete coefficients of 0, 1, and 2. During training label of each utterance is given to the model using the severity embedding that is added as an input to the variance adaptor before the pitch/energy/duration predictors. This allows the system to detect the relative characteristics of different severity groups, especially duration, pause and voice harshness, and variance of pitch and energy. It also allows additional control of the duration of the speech like the duration, pitch, and energy predictors, the dysarthria severity level predictor has a similar model structure which consists of a 2-layer 1D-convolutional network with ReLU activation, each followed by the layer normalization and a dropout layer, and an extra linear layer to project the hidden states into the output sequence (Ren, Hu et al. 2020).

### 2.3. Pause Insertion

Pause is another important indicator of dysarthric speech. The effect of sentence length on pause duration has also been previously investigated in persons with dysarthria due to Amyotrophic Lateral Sclerosis (ALS) (Rudzicz, Namasivayam et al. 2012). Their results show that the pause duration over sentence length for the group with higher severity level is increased by a higher rate in comparison with the group with lower severity level.

Analysis of the TORGO data set shows that the average Normal category speaker has only 0.26 between-word pauses (approximately only one such pause per every 4 complete number of between-word pauses of utterances), where a pause is defined as (Soleymanpour, Johnson et al. 2021). In comparison, the Low Dysarthria category has an average of 0.84 between-word pauses per utterance, 3.24 times higher than Normal, and the Moderate Dysarthria category has an average of 2.51 between-word pauses per utterance, 9.65 times higher than Normal. This is consistent with the prior literature for ALS patients.

To replicate the impact of pause on perception of dysarthric speech, it is straightforward to build a between-word pause model that simulates such pause characteristics. Although FastSpeech2 can already synthesize normal pause patterns for a given text, it is not sufficient to represent the patterns in dysarthric speech. To address this issue, we add a binary parameter to control insertion of additional pauses. Although pauses in dysarthric speech sometimes occur between phonemes within a word as well, the current version supports insertion of pauses only between words. To implement this, possible inter-word positions are identified, and then the maximum number of pauses is determined based on the severity level and length of the given sentence. For longer texts or for speakers with a higher dysarthria severity level, the model inserts more pauses. Since many of the sentences in the TORGO dataset are relatively short, there is not enough data to learn a complex model for pause insertion, so a simple model is used. The model uses the number of words in the sentence and the dysarthric severity level to determine the number of pauses to be inserted based on the average number of pauses in the TORGO data for that severity level. Once this is set, the locations of the pauses are chosen randomly at inter-word locations in the sentence. The pause insertion model is shown in the bottom left of the architecture in Figure 1.

For pause insertion, we aim to build a system to learn the pause patterns including both duration and frequency of pauses as the two main factors for each given speaker. To learn the pause length, the model considers this intrinsically as it would with other phonemes; thus, pause length is learned during training which is dependent on each speaker. Thus, for longer texts or for speakers with a higher dysarthria severity level, the model inserts more pauses. Full details of the pause insertion model are available in (Soleymanpour, Johnson et al. 2021).

*2.4. 5.1.3.   Augmentation using frame and phoneme level masking*

There are two options for pitch and energy modifications in the Variance adaptor shown in Figures 1 and 2, at both the phoneme and frame levels. In the frame level, the target duration is applied and then pitch and energy modifications are implemented while the modification of pitch and energy are carried out before the adjusting the target mel spectrogram duration in the phoneme level. The mel mask is used in the frame level modification, the source mask is applied in the phoneme level modification to modify pitch and energy. Masking is a method of padding to the maximum length of the input sequence which are phonemes or the maximum length of the output sequence which is here the mel spectrogram length.

The main paper (Ren, Hu et al. 2020) on which this approach is based was implemented based on frame level features for pitch and energy modifications. However, a recent variant of the paper found that phoneme level feature is more effective and their synthesized speech is more natural (Chien, Lin et al. 2021), so this approach is used here.

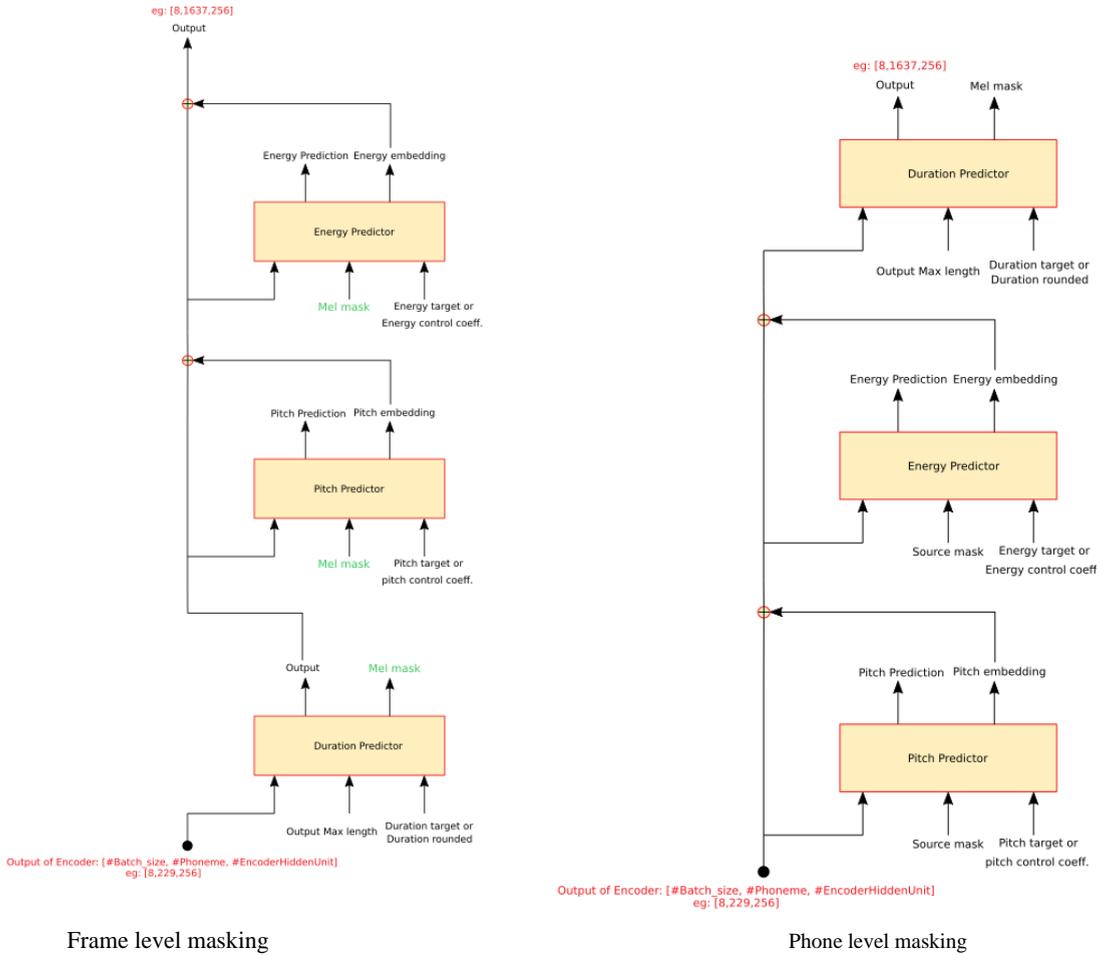

Frame level masking          Phone level masking

Figure 2- Variance adaptor architecture, a) Frame level Masking b) Phone level masking

## 3. Experimental setup

FastSpeech2 contains 4 feed-forward transformer blocks in the encoder and mel-spectrogram decoder. The decoder generates an 80-dimensional mel-spectrogram from hidden state. The size of the phoneme embedding is 256 in our implementation. The adjusted model was trained with a GeForce RTX 2080 Ti on the TORGO (Rudzicz, Namasivayam et al. 2012) dataset, containing 8 dysarthric speakers and 7 normal speakers.

After training the TTS models, the text in TORGO was used to synthesize additional dysarthric speech. The effect of the synthesized speech was evaluated by implementing two experiments on speech recognition application.

In the first experiment, the focus is on the effect of the severity and pause insertion level. Synthesized speech for augmentation was synthesized with three different severity coefficients of 0.0, 1.0, and 2, with the pause insertion turned on. Pitch, energy and duration coefficients were fixed at 1.0, the middle point within the allowable range. The number of augmented sentences was three times that of the original TORGO dataset.

For the second experiment, a wider range of dysarthric speech was synthesized for augmentation across all controllable parameters. Parameters for pitch, energy, duration, as well as severity level were varied across a range with pause insertion activated as shown in Table 5.1 below. The number of augmented sentences was ten times that of the original TORGO dataset.

Table 2- The prosody coefficients for synthesizing dysarthric speech in the two experiments

| Coef. | Baseline | Exp. 1 | Exp. 2 |
|---|---|---|---|
| Pitch | - | 1.0 | [0.1, 0.6, 1.2, 1.75] |
| Energy | - | 1.0 | [0.1, 1.0, 2.0] |
| Duration | - | 1.0 | [ 1.0, 1.3, 1.6, 1.8] |
| Severity level | - | [0.0, 1.0, 2.0] | [0.0, 1.0, 2.0] |
| Pause insertion | - | Yes | Yes |
| Total utterance | ~ 16000 | ~ ×3 | ~ ×10 |

The synthesized speech was used for training the DNN-HMM ASR model, trained on fMLLR transformed features Baseline configuration files provided in the Pytorch-kaldi repository for common speech databases like TIMIT, Librispeech were used as reference and the final architecture was based on experimental results using a small number of training set speakers (Khanal, Johnson et al. 2021). Our ASR models includes light bidirectional GRU (Ravanelli, Parcollet et al. 2019) architecture,  with five layers containing 1024 cells each, activated by Relu activation function and dropout of 0.2. The number of epochs was 10 to 12 to achieve the best result of each experiment. The architecture applies monophone regularization (Ravanelli, Brakel et al. 2018). A multi-task learning procedure was applied using two SoftMax classifiers, one estimating context-dependent states and the second one predicting monophone targets (Khanal, Johnson et al. 2021). For testing, a leave-one-speaker-out cross-validation procedure was applied across the original TORGO dataset.  Improvement to ASR accuracy was used to evaluate the effectiveness of augmenting the training data with synthetic dysarthric speech.

In addition, two subjective evaluation metrics were measured by trained human listeners. The first of these is the "dysarthric-ness" quality of the synthesized speech, which is compared for typical and dysarthric speakers when changing the severity level coefficients. The term "dysarthric-ness" is used to refer to the authenticity/accuracy of the synthesis engine in generating speech that sounds genuinely dysarthric to a human listener. Comparing the dysarthric-ness quality of synthesized speech for dysarthric and typical speakers shows us that the synthesized speech for dysarthric speakers is more naturally similar to real dysarthric speech than the speech synthesized using a typical non-dysarthric speaker as the synthesis target.

The second subjective evaluation metric is speaker similarity, a subjective metric of whether a human listener perceives that the synthesized dysarthric speech sounds like it has the same identity as the target speaker. This was used to evaluate the accuracy of utilizing specific speaker targets with only a very small amount of training data.

## 4. Result and discussion

*4.1. Observations*

Before evaluating the performance of the synthetic data augmentation on dysarthria-specific DNN-HMM speech recognition, we first observe and discuss the characteristics of the synthesized dysarthric speech.  Figure 3 shows the synthesized speech of speaker MC04 for the input text "We would like to play volleyball" for severity levels of 0 (Normal), 1 (Low Dysarthria) and 2 (Moderate Dysarthria), respectively. Pitch, energy and duration coefficients are the same across the various severity levels shown here. From the number of frames shown on the horizontal axis for each of the individual utterance segments, it is clear that the duration of the synthesized speech, one of the key indicators of dysarthria, increases with increasing severity level, independently of the separately controlled duration coefficient. Although Figure 3 shows the results for one specific speaker, this general pattern is true for all dysarthric speakers. However, the exact amount of impact on duration varies across speakers and phonemes, because it is learned from each speaker's individual speech characteristics and utterances.

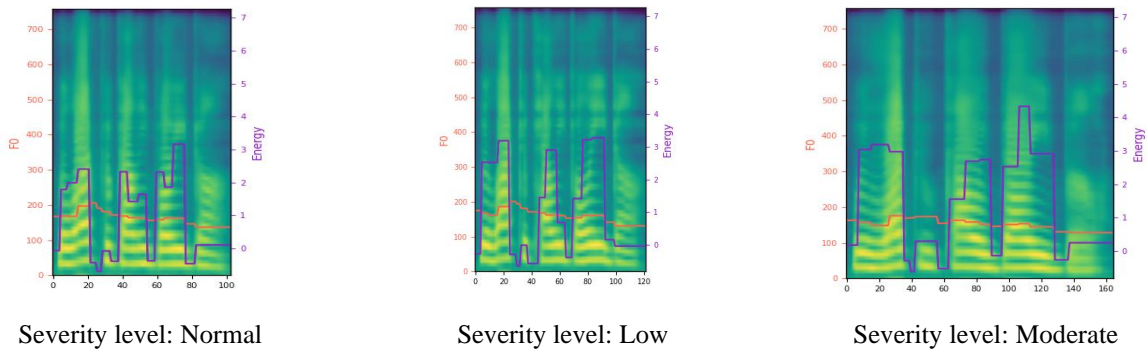

Severity level: Normal  Severity level: Low  Severity level: Moderate

Figure 3 Synthesized speech for Speaker MC04 "We like to play volleyball". Note the effect of the increasing dysarthria severity on the duration of the individual phonemes. These impacts are unique to each trained speaker.

To evaluate the effect of other parameters like pitch, energy and duration controllability, examples of synthesized speech for target speaker M05 with the input text "Bad and good" are presented in Figure 4. In this figure, the first row shows the results for increasing the Duration Coefficient from 1.0 to 1.3 to 1.6, with other factors fixed. The purple line indicates the energy envelope from which duration can be seen temporally. Based on the total number of frames it is clear that there is an increase in the total duration.

The second row of Figure 4 shows the spectrogram from increasing the energy coefficient from 0.5 to 1.0 to 2.0, with other factors fixed. The purple line shows the energy, corresponding to the right-hand vertical axis, indicating that an overall amplitude level corresponding to the energy coefficient is learned by the model from the training data.

The third row in Figure 4 plots the synthesized speech for the three pitch coefficients 0.5, 1.0, and 2.0. The orange line in these spectrograms shows the fundamental frequency, corresponding to the left-hand vertical axis, indicating that an increase in pitch corresponding to the pitch coefficient is learned by the model from the training data.

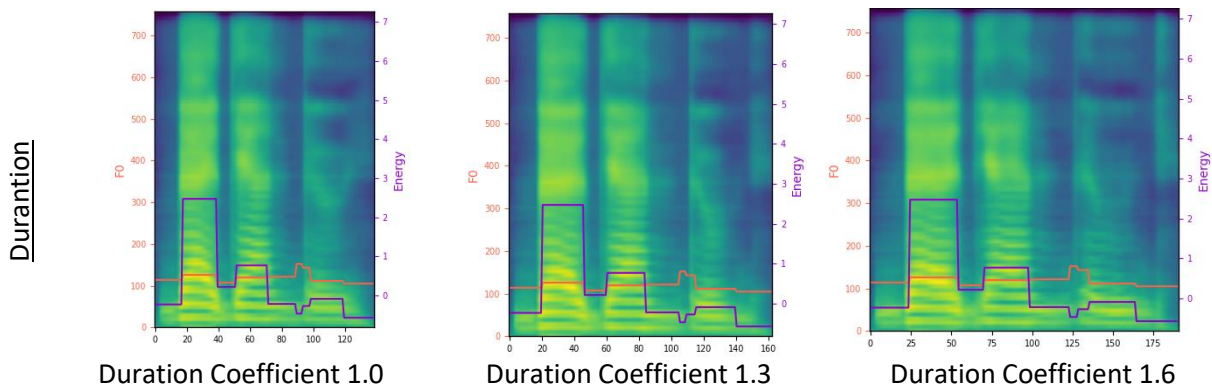

Duration Coefficient 1.0  Duration Coefficient 1.3  Duration Coefficient 1.6

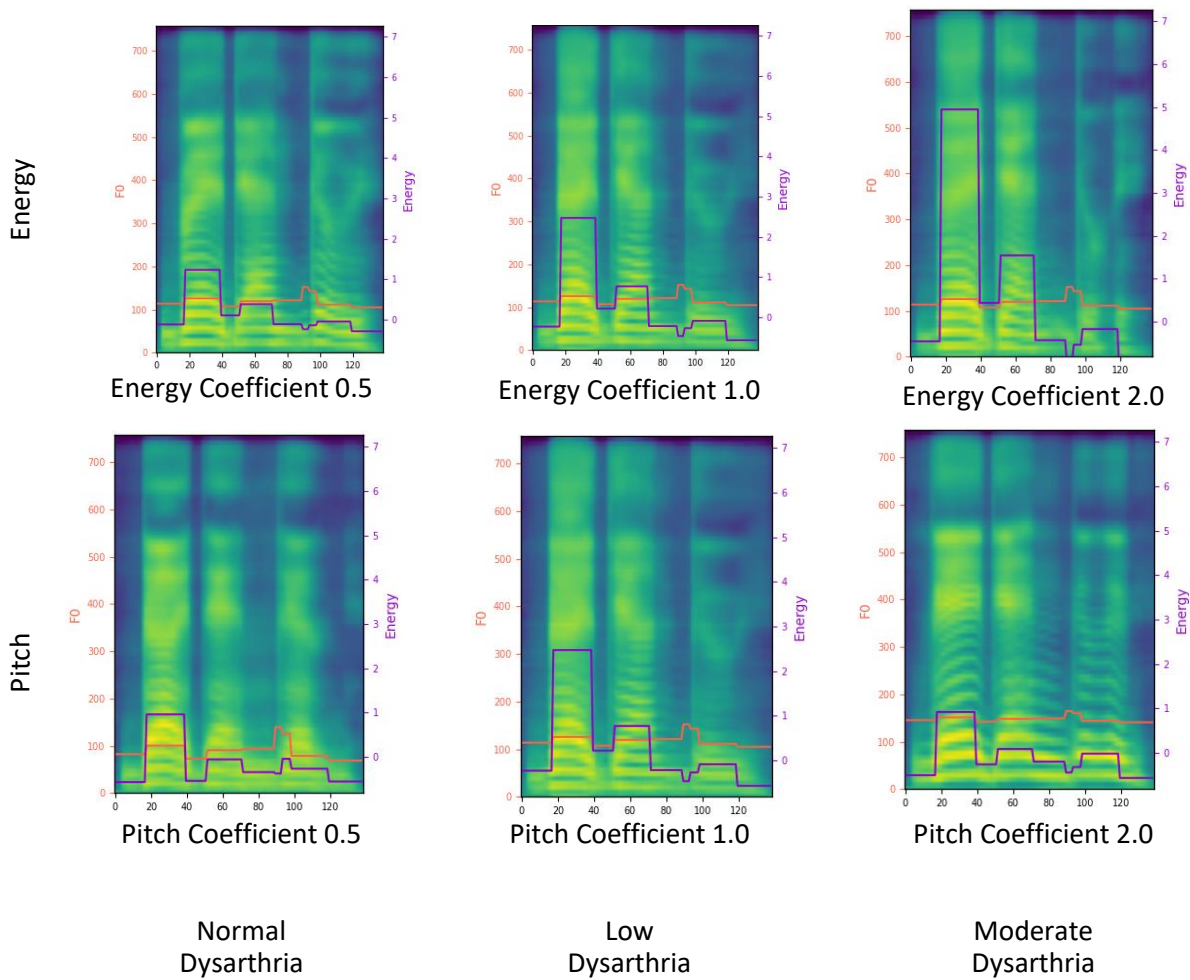

Figure 4- Synthesized speech for Speaker M05 "Bad and Good". Rows correspond to varying one specific prosodic coefficient while leaving the others constant.

In addition to these explicit characteristics controlled by specific coefficients, many individual dysarthric characteristics are learned by the system itself from the training data corresponding to individual speakers, as part of the speaker embedding process that models the target speaker. Specific examples of this include:

- The system learned to model stutter-like characteristics for some speakers. For example, synthesized speech for the sentence "How can we synthesize better dysarthric speech?", trained on target speaker M05, resulted in a stutter occurring just before the phone /b/ in the word "best". There is no explicit stutter parameter in the model, but this prosodic characteristic was learned through training data. You can listen to the audio in the demo page1.
- The system learned specific phoneme confusion patterns around place and manner of articulation. For example, synthesized speech for "This is the pad", trained on target speaker MC02, resulted in the 11phoneme /p/ being synthesized in a manner perceived by listeners as the phoneme /c/. Note also that this pronunciation confusion was co-located with an increase in duration representative of the common pattern of extended duration of the phoneme /æ/ in words such as bad. Again, there is no explicit parameter corresponding to place and/or manner of articulation, with this characteristic being learned by the synthesis model through training data.

---

[1] https://mohammadelc.github.io/SpeechGroupUKY/#:~:text=Bad%20and%20good%22-,Other%20Observation,-We%20have%20noticed

- The system learned between-word pause durations corresponding to individual speaker patterns. Even though the system has a specific controllable parameter for pause insertion frequency, the duration of these pauses is learned implicitly by the synthesis model for each individual speaker. As an example, the synthesized speech for the sentence "How can we synthesize better dysarthric speech?" trained on speakers M02 and M05 shows much longer inserted pauses for M05 compared to M02, given identical pause insertion parameters.

## 4.2. Objective evaluation using ASR Word Error Rate

To evaluate the results of the two data augmentation ASR experiments as mentioned Section 5.2, the Word Error Rate (WER) was calculated for each test speaker, with varying amounts and types of training data coming from the synthesized speech for those target speakers. As discussed above, the first experiment focuses on speech augmentation using severity level and pause insertion, with data augmentation three times the size of the original dataset. The second experiment includes a wide range of controllable parameters for pitch, energy, duration, severity level, and pitch insertion, with data augmentation ten times the size of the original dataset.

Table 5.2 shows the WER of the two experiments along with the baseline and compares them with the results of the best models of two other published works performed on the same TORGO dataset using hybrid speech recognition models (Espana-Bonet and Fonollosa 2016, Yue, Christensen et al. 2020).

Results show that the WER performance of the baseline is similar to that of the two comparison methods for the lowest few severity levels, and slightly better for the highest ("medium") severity. The average WER across all speakers is 44.5%, 56.2% and 43.3% for our baseline, (Yue, Christensen et al. 2020) and (Espana-Bonet and Fonollosa 2016), respectively.

In the first experiment that only used severity synthesis and pause insertion, the synthesized speech used for augmenting ASR training improved the performance of the DNN-HMM model for each speaker except M03, which declined slightly. Average WER performance across all speakers improves, with WER dropping from 44.5% to 41.6%. The second experiment with additional prosody variance and data augmentation shows further performance improvement, with individual improvement for all 8 speakers in the dataset.

Average WER performance across all speakers improves, with WER dropping from 44.5% to 41.6% (first experiment) to 39.2% (second experiment). Relative to the baseline, this corresponds to a reduction in WER by 12.2% due to the data augmentation.

Table 3-WER of each test speaker for the two augmentation experiments: Exp.1 included augmented speech across 3 severities with pause insertion (augmented data 3x original), and Exp. 2 included augmented speech across severity, pause, pitch, energy, and duration (augmented data 10x original)

| Severity Level | Test Spk | WER (%) | | | | |
|---|---|---|---|---|---|---|
| | | Baseline | Exp. 1 | Exp. 2 | (Yue, Christensen et al. 2020) | (Espana-Bonet and Fonollosa 2016) |
| Very low | F04 | 16.8 | 16.3 | 14.5 | 18.3 | 13.1 |
| | M03 | 10.9 | 12.7 | 10.7 | 18.2 | 17.7 |
| Low | F03 | 46.6 | 39.3 | 36.8 | 44.2 | 39.1 |
| Moderate | F01 | 58.3 | 52.4 | 50.4 | 71.5 | 39.6 |

|  | M01 | 55.4 | 51.3 | 50.3 | 69.3 | 62.2 |
|---|---|---|---|---|---|---|
|  | M02 | 44 | 43.1 | 38.4 | 70.9 | 42.9 |
|  | M04 | 65.8 | 64.2 | 62 | 79.9 | 69.0 |
|  | M05 | 58.2 | 53.6 | 49.6 | 77.2 | 62.6 |
| **Overall Average** |  | 44.5 | 41.6 | 39.2 | 56.2 | 43.3 |

To separate the effect of the proposed approaches as a function of the level of severity of the dysarthric speech, Table 5.3 shows the average WER for speakers at the different dysarthria severity levels. This shows that some augmentation using synthetic speech at three dysarthria levels with pause insertion improves the WER of each severity level on average except for the group with the low severity. A higher level of augmentation using synthetic speech across all controllable parameters improved WER across all severity levels. The biggest improvement was in the Low Dysarthria category, with a relative reduction in WER of 21% compared to the baseline.

Table 4- WER of each severity level for the two augmentation experiments.

| Severity level | Baseline | Exp. 1 | Exp. 2 | Decrease in WER Relative to Baseline | |
|---|---|---|---|---|---|
|  |  |  |  | Exp. 1 | Exp. 2 |
| **Very Low** | 13.8 | 14.5 | 12.6 | -4.7% | 9% |
| **Low** | 46.6 | 39.3 | 36.8 | **7.3%** | **21%** |
| **Moderate** | 56.3 | 52.9 | 50.1 | **6%** | **11%** |
| **All** | 44.5 | 41.6 | 39.2 | **6.5%** | **12.2%** |

### 4.3. Subjective evaluation

In this section, we present a subjective evaluation for the two dimensions of dysarthric-ness and speaker similarity. As discussed under Experimental Setup above, dysarthric-ness is a measure of whether the synthetic speech sounds genuinely dysarthric to a human listener, and speaker similarity is a measure of whether the synthetic speech sounds like it has the same identity as the target speaker to a human listener. These metrics are similar to what has been used in the speech community to evaluate Voice Conversion (VC) algorithms, only here we apply them to dysarthric synthesis.

To implement this evaluation we asked individuals with appropriate speech pathology clinical experience to conduct the evaluations of the synthesized dysarthric speech.

For the dysarthric-ness measure, three evaluators were asked to rate 45 utterances, 10 examples of synthesized dysarthric speech interlaced with 5 examples of actual dysarthric speech as control examples, for each of 3 target speakers representing 3 different severity levels. The prompt for evaluators is as follows.

> "Listen to the following audio and assess whether the speech sample sounds more like Dysarthric (e.g. from an ALS patient) or Healthy speech. Focus specifically on those characteristics that you associate with dysarthric speech, as opposed to the pronunciation, speech quality, or presence of processing artifacts. Please evaluate the degree of perceived dysarthria on the 4 point scale: "
> *Definitely Healthy | Probably Healthy | Probably Dysarthric | Definitely Dysarthric*.

For the speaker similarity measure, three evaluators were asked to rate 40 utterance pairs. This included 10 examples comparing synthetic speech of a target speaker to actual speech of that same utterance from the same original speaker, and 10 examples comparing the synthetic speech of one target speaker to actual speech of that same utterance spoken by a different speaker having the same dysarthria level. This was done for each of two target speakers, one of which was very low dysarthria (M03) and one of which was moderate dysarthria (M02). The prompt for the evaluators is as follows:

> "Do you think these two samples could have been produced by the same speaker? Some of the samples may sound somewhat degraded/distorted. Please try to listen beyond the distortion and concentrate on identifying the voice. Are the two voices the same or different? You have the option to indicate how sure you are of your decision." Participants were asked to rate the stimuli on a 4-point scale:

*Definitely the Same | Probably the Same | Probably Different | Definitely Different.*

**Dysarthric-ness Results**

Figure 5 shows the results of dysarthric-ness for the three different dysarthria severity levels. As mentioned before, we included some real dysarthric speech to reduce the bias effect of the raters. The four categories of Definitely Healthy, Probably Healthy, Probably Dysarthric, and Definitely Dysarthric were assigned numerical values of 1, 2, 3, and 4 in order to compute the "average" perceived dysarthric-ness for each experiment.

For the Very Low severity group, the real speech was entirely perceived as Definitely (50%) or Probably (50%) Healthy, with an average perceived dysarthric-ness of 1.50. In comparison, the synthesized speech was perceived as more dysarthric, with the four categories from Definitely Healthy to Definitely Dysarthric evaluated at 0%, 44.4%, 38.9%, and 16.7%, giving an average perceived dysarthric-ness of 2.72. This indicates that our synthesized speech was not perceived very accurately for Very Low severity.

For the Low severity group, the real speech was perceived across the four categories from Definitely Healthy to Definitely Dysarthric as 11.1%, 50%, 38.9%, and 0%, for an average dysarthricness of 2.28. The synthesized speech was perceived as slightly more healthy, with the four categories as 26.7%, 13.3%, 33.3%, and 26.7%, for an average dysarthricness of 2.60. It is interesting that for the Low category, the synthesized speech was perceived as more healthy rather than more dysarthric, in comparison to the Very Low category.

For the Moderate severity group, the highest severity level considered, the real speech was perceived across the four categories from Definitely Healthy to Definitely Dysarthric as 0%, 22.2%, 36.1%, and 41.7%, for an average dysarthricness of 3.19. The synthesized speech was perceived as slightly more healthy, with the four categories as 2.8%, 19.4%, 61.1%, and 16.7%, for an average dysarthricness of 2.92. This group has the closest ratings, suggesting that the perceived dysarthricness becomes more similar between synthesized and real speech as the dysarthria severity level increases. As with the Low severity group, the Moderate group synthesized speech was perceived as slightly healthier than the actual speech.

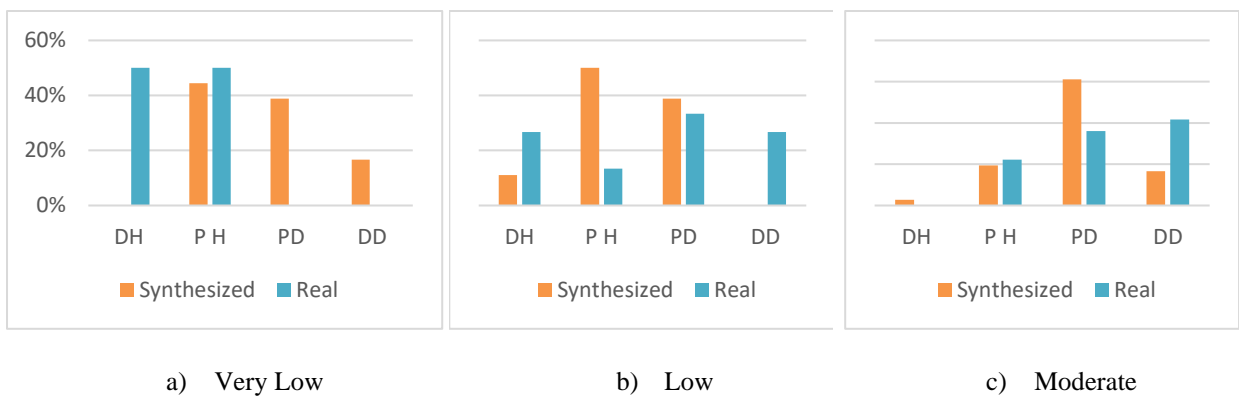

Figure 5- Perceived dysarthric-ness at different severity levels, for synthesized speech compared to real speech

Another interesting observation relates to the degree of certainty of evaluations, by considering Definitely vs. Probably evaluations. For the Very Low category, real speech evaluations were 50% in the "Definitely" compared to only 16.7% of the synthesized speech evaluations, so there was a higher certainty for evaluating real speech. For the Low category, real speech evaluates were 53.3% "Definitely" compared to 11.1% for synthesized speech evaluations, and for the Moderate category, real speech evaluations were 41.7% "Definitely" compared to 19.4% for synthesized speech evaluations. Across all three severities, real speech evaluations were 48.3% "Definitely" versus synthesized speech evaluations at only 15.7% definitely. This suggests that evaluators found the task more challenging for the synthesized speech.

Overall, the perceived dysarthricness of the synthesized speech tracks with the actual dysarthricness of the trained target speaker, becoming closer in terms of perceived dysarthricness at higher levels of dysarthria severity. Figure 6 below shows the overall perceived dysarthricness compared between real and synthesized speech across all severities.

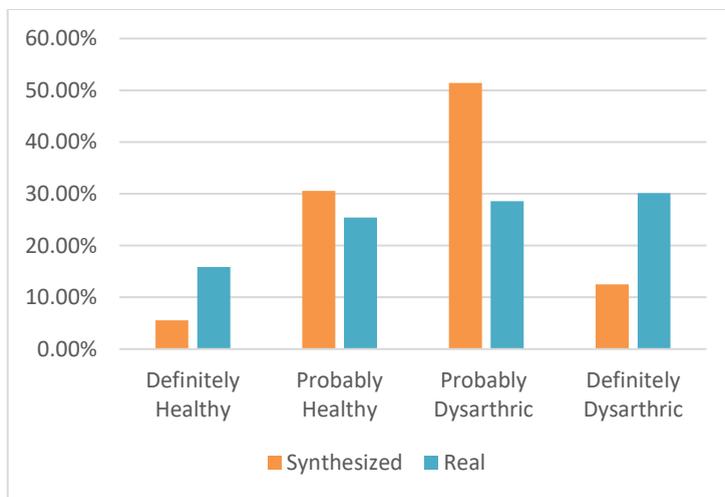

Figure 6. Overall Dysarthric-ness of all groups combined

**Speaker Similarity Results**

Table 4 show the results of the Speaker Similarity evaluation. As would be expected, the judged speaker similarity is higher when the target speaker of the synthesized speech is the same as the speaker of the real speech. Interestingly, for both the same and different target speakers, synthetic speech for a target speaker with a very low level of dysarthria was judged to have lower speaker similarity.

Table 4. Speaker Similarity subjective evaluation results

| Same / Diff Speaker | Dysarthria of real speech | Definitely Same | Probably Same | Probably Different | Definitely Different | Weighted Speaker Similarity (4.0 scale) |
|---|---|---|---|---|---|---|
| **Different** | Very Low | 3.3% | 3.3% | 40.0% | 53.3% | 1.57 |
| **Different** | Moderate | 3.3% | 23.3% | 30.0% | 43.3% | 1.87 |
| **Same** | Very Low | 0.0% | 33.3% | 43.3% | 23.3% | 2.10 |
| **Same** | Moderate | 13.3% | 43.3% | 26.7% | 16.7% | 2.53 |

Overall, the speaker similarity rating for the synthesized speech when the target speaker matches the real speech seems fairly low, averaging only 2.32 on a 4.0 scale with only 13.3 % of evaluations as "definitely the same" speaker. This suggests that the aspects of the speech synthesizer relating to speaker characteristics still has significant room for improvement, although it is also significantly impacted by the very low amount of data from each individual speaker available for training the models. This overall result is shown in Figure 7 which presents a bar chart of the 4 evaluation categories for the cases when the synthesized speaker is different than actual and when the synthesized speaker is the same as actual.

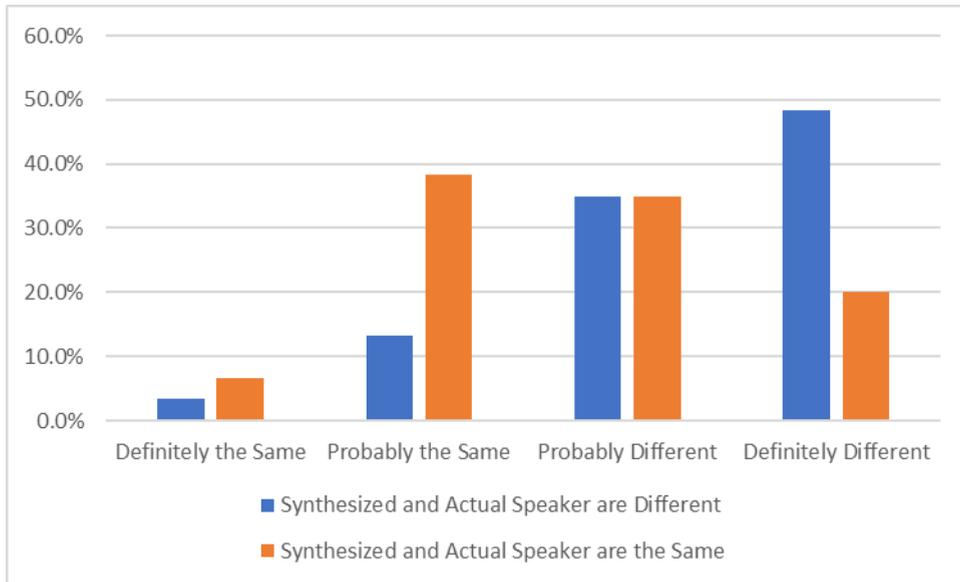

Figure 7. Speaker Similarity rating between synthetic and real speech, when speakers are the same compared to when speakers are different

As with the Dysarthricness evaluation, an estimate of degree of certainty can also be obtained by comparing the by considering Definitely vs. Probably evaluations. For Speaker Similarity, when the synthesized and actual speaker are different, raters were significantly more confident, with 51.6% of the evaluations being definite in their evaluations (48.3% rating "definitely different" and 3.3% rating "definitely the same"). In contrast, when the synthesized and actual speaker are the same, raters were much less confident, with only 26.7% being definite in their evaluations (20.0% rating "definitely different" and 6.7% rating "definitely the same"). This suggests evaluators found it challenging to decide whether actual speech and synthetic speech for that same target speaker were representative of the same speaker.

## 5. Conclusion

In this paper, we have modified a neural multi-talker TTS by adding a dysarthria severity level coefficient and a pause insertion model to synthesize dysarthric speech for varying severity levels, for the purpose of providing data augmentation for machine learning tasks such as automatic speech recognition. We evaluate the effectiveness of the approach for dysarthria-specific speech recognition on the TORGO dataset, with results provided for two different experiments: the first includes augmented speech across 3 severities with pause insertion, and the second includes augmented speech with across a larger number of variables that include severity, pause, pitch, energy, and duration.

Overall results show a relative improvement of 12.2% on Word Error Rate (WER), demonstrating that using dysarthric synthetic speech to increase the amount of dysarthric-patterned speech for training has the potential for significant impact on dysarthric ASR systems. In addition, two subjective evaluations of the synthesized dysarthric speech are provided. This includes an evaluation of Dysarthric-ness that shows that the perceived level of the dysarthria tracks with the target synthesized dysarthric severity, as well as an evaluation of Speaker Similarity that shows a higher rating of similarity between synthesis target speaker and actual speaker when these are the same individual, although the amount of data available data for training is not enough to fully capture the voice characteristics of each speaker.

A demonstration web page with audio results of the synthesis is available at
https://mohammadelc.github.io/SpeechGroupUKY/.

## Acknowledgements

This work was supported by National Institutes of Health under NIDCD R15 DC017296-01.